\documentclass[aps,prb,reprint,groupedaddress,citeautoscript,longbibliography,twocolumn,amsmath]{revtex4-2}
\pdfoutput=1
\usepackage{graphicx}
\usepackage{dcolumn}
\usepackage{bm}
\usepackage{bbold} 
\usepackage[T1]{fontenc}
\usepackage[]{threeparttable}
\newcommand{\bra}[1]{\langle #1 | }
\newcommand{\ket}[1]{| #1 \rangle}
\begin{document}

\preprint{APS/123-QED}

\title{Hole Spin in Direct Bandgap Germanium-Tin Quantum Dot}

\author{Nicolas Rotaru}
\author{Patrick Del Vecchio}%
\author{Oussama Moutanabbir}
    \affiliation{Department of Engineering Physics, \'Ecole Polytechnique de Montr\'eal, Montr\'eal, C.P. 6079, Succ. Centre-Ville, Montr\'eal, Qu\'ebec, Canada H3C 3A7}
\date{\today}

\begin{abstract}
Germanium (Ge) has emerged as a contender for scalable solid-state spin qubits. This interest stems from the numerous attractive properties of hole spin in Ge low-dimensional systems and their compatibility with the standards of silicon processing. Herein, we show that the controlled incorporation of Sn into the Ge lattice enables hole spin quantum dots that retain the same advantages as those made of Ge while also providing bandgap directness. The latter is essential for a more efficient interaction with light, a key feature in the implementation of photon-spin interfaces and quantum memories. We first map the material properties for a range of Ge$_{1-x}$Sn$_x$ planar heterostructures to identify the optimal conditions to simultaneously achieve hole spin confinement and bandgap directness.
Although compressive strain is necessary for heavy hole confinement, we estimate that an additional 4.5 at.\% of Sn is needed for every 1\% increase in the absolute value of compressive strain to preserve the direct bandgap.  However, a high compressive strain is found to be detrimental to the Rashba coupling.  Moreover, a theoretical framework is derived to evaluate the dipole moment $d$ and the relaxation rate $\Gamma$ of electric dipole spin resonance quantum dot devices. We compare the perturbative and effective values of $d$ with the values obtained from the full 3D Hamiltonian. We find $d$ to be around 1 and 0.01 e pm for the out-of-plane and in-plane configurations, respectively, and $\Gamma\propto B^5$, eventually becoming $\propto B^7$ in the out-of-plane configuration.  

\end{abstract}

\maketitle

Hole spin solid-state devices have recently received great interest as a reliable building block for quantum processors and simulators \cite{burkard2023semiconductor}. In particular, hole spin in germanium (Ge) has been one of the most attractive spin systems as it is associated with several advantages \cite{scappucci2021,jirovec2021singlet,watzinger2018germanium,liu2022gate,liu2023ultrafast,DelVecchio_2023,borsoi2024shared,van2024coherent,mironov2014ultra,philippopoulos2020first,goswami2007controllable,moutanabbir2024nuclear}: 
(i) high hole mobility;
(ii) strong spin-orbit coupling (SOC);
(iii) strain-induced tunable coupling between light-hole (LH) and heavy-hole (HH) bands;
(iv) \textit{p}-symmetry of the valence band associated with reduced hyperfine interactions;
(v) absence of valley degeneracy, a major challenge for electrons in Si;
(vi) high natural abundance of nuclear-spin free isotopes, which further suppresses the nuclear spin bath decoherence channel.
Using isotope purification, the nonzero nuclear spin isotope content can be brought below 0.01\% in quantum wells (QWs)  and practically eliminated in the volume of an electrostatically defined quantum dots (QDs) \cite{moutanabbir2024nuclear}.
However, Ge is an indirect bandgap material, which
can limit the potential introduction of Ge hole spin qubits in quantum systems that require the conversion of a stationary carrier quantum state to a flying optical qubit, and \textit{vice versa}, because absorption and emission are weak in an indirect bandgap material. These limitations can be alleviated in a direct bandgap material, allowing additional
flexibility in the
design of a variety of photon-spin platforms such as quantum repeaters, a necessary component to implement, for instance, long-distance quantum networks \cite{gaudreau2017entanglement}. 

Herein, we propose an all-group IV system consisting of a silicon-integrated Germanium-Tin alloy (Ge$_{1-x}$Sn$_{x}$) that exhibits all key attributes of the hole spin in Ge while also being of direct bandgap. 
When Ge is alloyed with $\alpha$-Sn, a  zero bandgap semimetal, the $\Gamma$ valley lowers more rapidly with the Sn content than the $L$ valley, eventually leading to
a direct bandgap semiconductor.
Owing to its tunability and compatibility with silicon, this alloy has been the subject of extensive studies toward the monolithic integration of photonics and electronics \cite{moutanabbir2021monolithic}. In addition, Ge$_{1-x}$Sn$_{x}$-based heterostructures have recently been introduced to achieve selective confinement of LH in Ge \cite{assali2022light,DelVecchio_2023,del2024light}, thus providing a range of opportunities to study spin physics and spin devices in group IV systems.

In the following, we start by investigating band lineups and effective parameters of HH in  Ge$_{1-x}$Sn$_x$ QWs and discuss the behavior of electrostatically defined QDs. We also study, as a control scheme, the electric-dipole spin resonance (EDSR), which allows the electrical control of the spin state by leveraging the large SOC \cite{rashba1991}. Finally, we estimate the phonon-assisted relaxation rate in Ge$_{1-x}$Sn$_{x}$ QDs. By mapping these parameters and the associated performance, this work lays the groundwork to introducing Ge$_{1-x}$Sn$_x$ in design of direct bandgap all-group IV spin qubits.

\section{Theoretical Framework}

\subsection{The 8 band $k\cdot p$ model}
To evaluate the material parameters needed to confine HHs  in Ge$_{1-x}$Sn$_{x}$ QWs, a $k\cdot p$ model is employed.
The 8-band $k\cdot p$ matrix $\mathbf{H}_{k\cdot p}(\mathbf{K})$ is written as \cite{bahder1990eight,Eissfeller2012}:
\begin{equation}
\label{eq:QWHamiltonian}
     \mathbf{H}_{k\cdot p}(\mathbf{K}) = \mathbf{H}_{k}(\mathbf{K}) + \mathbf{H}_{SO} + \mathbf{H}_{\epsilon} + \mathbf{H}_{B}+\mathbf{H}_q(\mathbf{K})
\end{equation}
Theses matrices, given in Appendix \ref{sec:kaneHamiltonian}, represent the effect of the hole kinetic energy within the lattice, the SOC, the strain, the external magnetic field, and the cubic in $\mathbf{J}$ correction to the g-factor, respectively \cite{Winkler2003}.
Growth along the [001] direction and bi-axial isotropic strain are considered. The strain in-plane components are obtained from lattice mismatch $\epsilon_\parallel= a/a_0-1$,
where $a$ is the in-plane lattice parameter (kept constant in the heterostructure) and $a_0$ is the relaxed lattice parameter.  With $z$ as the growth direction, the Hamiltonian of the QW, $\mathbf{H}_{\text{QW}}$, is written as:
\begin{equation}
    \mathbf{H}_{\mathrm{QW}} = \mathbf{H}_{k\cdot p}(\mathbf{K})+V(z),
\end{equation}
where $V(z)$ is the band alignment. This potential is the sum of the average valance band offset energy $E_{v,avg}$ \cite{van1989band} and the potential energy associated with the external out-of-plane component of the electric field $E_z$:
\begin{equation}
    V(z) =E_{v,avg}(z)+eE_z z
\end{equation}
The mechanical wave vector is $\mathbf{K}=\mathbf{k}+e\mathbf{A}/\hbar$, where $\mathbf{k} \rightarrow -i\nabla$ is the canonical wave vector and $\mathbf{A}$ the vector potential. In this work, three different gauges are employed, depending on the magnetic field orientation and in-plane confinement. For a QD in arbitrarily oriented magnetic fields:
\begin{align}
\begin{split}
\label{eq:gauge3D}
     \mathbf{A}(\theta,\phi) &=\frac{B}{2}\cos \theta \left[ -y \mathbf{e}_{x} +x\mathbf{e}_{y}\right] \\
     &-B\sin \theta (x\sin \phi -y\cos \phi )\mathbf{e}_{z}, \
     \end{split}
\end{align}
where $\theta$ and $\phi$ are, respectively, the polar and azimuthal angle of the applied magnetic field $\mathbf{B}=\nabla \times \mathbf{A}$. In the QW system (no in-plane confinement), the following two vector gauges are used in the case of an out-of-plane ($\mathbf{B}_\perp $) and in-plane ($\mathbf{B}_\| $) $\mathbf{B}$-fields, respectively \cite{del2024light}:
\begin{align}
        \mathbf{A}_\perp  &= B/2(-y \mathbf{e}_x+x \mathbf{e}_y)\label{eq:gaugeBperp}  
        \implies \mathbf{B}_\perp = B \mathbf{e}_z  \\
\begin{split}
        \mathbf{A}_\| &= z B (\sin \phi \mathbf{e}_x-\cos \phi \mathbf{e}_y) \label{eq:gaugeBparr} \\
        &\implies \mathbf{B}_\| =  B (\cos \phi \mathbf{e}_x +\sin \phi \mathbf{e}_y)
\end{split}
\end{align}
While Eq. (\ref{eq:gaugeBperp}) is just a special case of Eq. (\ref{eq:gauge3D}), it is necessary to use a gauge different from Eq. (\ref{eq:gaugeBparr}) for the in-plane case so that the effective parameters of the QW states under $\mathbf{B}_\parallel$ do not depend on $x$ and $y$.

\subsection{The Quantum Well Effective Hamiltonian}

Because of the lack of coupling between HH and other states at the $\Gamma $ point, the eigenstates of $\mathbf{H}_{\text{QW}}$ at $K_x=K_y=0$ are either pure HH ($H$ states) or a mix of LH, split off (SO), and conduction band (CB) states, which are labeled $\eta$ states:
\begin{align}
\label{eq:subband_enveloppes}
|H,\sigma \rangle &= |HH,\sigma\rangle | h \rangle \\
|\eta,\sigma \rangle &=|CB,\sigma \rangle |c\rangle +|LH,\sigma \rangle |l\rangle +\sigma |SO,\sigma \rangle |s\rangle
\end{align}

$H$ and $\eta$ states serve as the basis onto which the Hamiltonian, at finite $\mathbf{K}$ and $\mathbf{B}$, is projected. The result of this projection using the general gauge in Eq. (\ref{eq:gauge3D}) is given in Appendix \ref{sec:subbandedgebasis}.
A Shrieffer-Wolff transformation (SWT) is then performed on the projected Hamiltonian to  obtain the effective Hamiltonian of a given subband, and treat other subbands perturbatively \cite{Winkler2003,DelVecchio_2023}. Keeping only terms that are exact up to $\mathbf{K}^3$ yields:

\begin{align}
\mathbf{H}^{\mathrm{eff}}_{\perp}  &=\alpha_0 \gamma K_{\|}^2 
+\frac{\alpha_0}{l_B^2} \frac{g^{\mathrm{QW}}_{\perp}}{2} \sigma_z + \mathbf{H}_{\mathrm{R}}(\mathbf{K}) \label{eq:effHperp} \\
\mathbf{H}^{\mathrm{eff}}_{\|}  &=\alpha_0 \gamma k_{\|}^2
+\frac{\alpha_0}{l_B^2} \frac{g^{\mathrm{QW}}_{\|}}{2} (e^{-i\phi}\sigma_+ + \mathrm{h.c.}) +\mathbf{H}_{\mathrm{R}}(\mathbf{k})
\label{eq:effHparr}
\end{align}
where $\alpha_0=\hbar^2/2m_0$, $m_0$ is the free electron mass, $\hbar$ is the reduced Planck constant,
$l_B=\sqrt{\hbar/eB}$ is the magnetic length, 
$\sigma_{\pm}  = (\sigma_x \pm i \sigma_y)/2$, and $\sigma_{x,y,z}$ are the Pauli matrices.
The terms in Eq. (\ref{eq:effHperp}) represent respectively the parabolic dispersion (with effective mass $\tilde{m} = m_0/\gamma$), the linear Zeeman splitting (with effective perpendicular g-factor $g^{\mathrm{QW}}_\perp$), and the Rashba coupling term: 
\begin{equation}
\begin{split}
   \mathbf{H}_\mathrm{R} =  & i\beta_1(k_{-}\sigma_--k_{+}\sigma_+)  
 -i \beta_2\left(k_{-}^3 \sigma_{+}-k_{+}^3 \sigma_{-}\right) \\
& +i \beta_3\left(k_{-} k_{+} k_{-} \sigma_{-}-k_{+} k_{-} k_{+} \sigma_{+}\right),
\end{split}
\end{equation}
that includes the linear Rashba coupling ($\beta_1$), and the two cubic Rashba coupling ($\beta_2$, $\beta_3$), with  $k_\pm = k_x \pm i k_y$.
The second term of Eq. (\ref{eq:effHparr}) contains the effective parallel g-factor term, $g^{\mathrm{QW}}_\|$. In this equation, $\mathbf{k}$ is canonical since the field is in-plane and we have integrated over $z$ while projecting on the \{$H$,$\eta$\} basis. The expressions for the effective parameters are given in Appendix \ref{sec:EffectiveParameters}.

\subsection{The Quantum Dot Hamiltonian}
The Hamiltonian of the QD is defined by adding an in-plane parabolic confinement $V(x^2,y^2)$ of effective length $l_x$ and $l_y$:
\begin{equation}
    \mathbf{H}_{\mathrm{QD}} =\mathbf{H}_{\mathrm{QW}}-\alpha _{0}\zeta\left(\frac{x^{2}}{l_{x}^{4}} +\frac{y^{2}}{l_{y}^{4}}\right)
    \label{eq:HamiltonianQD}
\end{equation}
where $\zeta$ is a scaling coefficient. The ladder operators associated with the in-plane potential are:
\begin{align}
a_{x} =\frac{1}{\sqrt{2}}\left(\frac{x}{l_{x}} +il_{x} k_{x}\right),
a_{y} =\frac{1}{\sqrt{2}}\left(\frac{y}{l_{y}} +il_{y} k_{y}\right), \label{eq:ladder}
\end{align}
which act on the in-plane states in the usual way; $a_{i} |n_{i}\rangle = \sqrt{n_{i}}|n_{i}-1\rangle$ and $a_{i}^\dagger |n_{i}\rangle  = \sqrt{n_{i}+1}|n_{i}+1\rangle$. 
The matrix representation of the ladder operators defined by Eq. (\ref{eq:ladder}) can be written up to a certain value of $n$ and used to write the in-plane terms of $\mathbf{H}_{\mathrm{QD}}$. The general gauge in Eq. (\ref{eq:gauge3D}) is used to allow for an arbitrary configuration of the magnetic field. 
An effective expression of the QD Hamiltonian, $\mathbf{H}_{\mathrm{QD}}^{\mathrm{eff}}$, can be written using the effective parameters of a given QW subband:
\begin{equation}
\label{eq:effQD}
    \mathbf{H}_{\mathrm{QD}}^{\mathrm{eff}} = \mathbf{H}_{\mathrm{FD}}+ \mathbf{H}_{\mathrm{R}},
\end{equation}
where $\mathbf{H}_{\mathrm{FD}}$ is the effective Hamiltonian of a symmetric QD ($l_x = l_y = l)$ neglecting the Rashba coupling terms:
\begin{equation}
    \mathbf{H}_{\mathrm{FD}}= \alpha_0 \gamma K_\|^2 + \frac{\alpha_0}{2 l_B^2} \boldsymbol{\sigma} \cdot \mathbf{g} \cdot \mathbf{n}  - \alpha_{0}\zeta\left(\frac{x^{2}}{l^{4}} +\frac{y^{2}}{l^{4}}\right),
\end{equation}
where $\boldsymbol{\sigma}$ is the vector of Pauli matrices, $\mathbf{n}$ is the direction of the magnetic field, and  the $\mathbf{g}$-matrix is:
\begin{equation}
    \mathbf{g} = 
    \begin{bmatrix}
g^{\mathrm{QW}}_{||} & 0 & 0\\
0 & g^{\mathrm{QW}}_{||} & 0\\
0 & 0 & g^{\mathrm{QW}}_{\perp }
\end{bmatrix}
\end{equation}
The solutions of $\mathbf{H}_{\mathrm{FD}}$ when $\mathbf{n}=\mathbf{e}_z$ and $\zeta=\gamma$ are the well-known Fock-Darwin orbitals.

\section{Results and discussion}
\subsection{Ge/Ge$_{1-x}$Sn$_x$/Ge band alignment}

The evolution of the $\Gamma$ and $L$ critical point is characterized by a linear interpolation between the values of the bulk materials and the respective bowing parameters $b_\Gamma$ and $b_L$:
\begin{equation}
    E_i(x) = (1-x)E^{\mathrm{Ge}}_i+(x) E^{\mathrm{Sn}}_i - b_i x(1-x),
\end{equation}
with $i={\Gamma,L}$. These parameters have been reported in both experimental and theoretical studies suggesting a 1.9-3.1 eV range for the direct gap and 0.26-1.23 eV  range for the indirect one \cite{mkaczko2016material,d2006optical}.
Some studies reported better agreement with experiments for composition-dependent bowing \cite{gallagher2014compositional}, while others claim that, above a certain Sn content, a composition-independent bowing is more accurate \cite{d2016near}.
For this study, $b_\Gamma$=2.46 eV \cite{bertrand2019experimental} and $b_L$=1.23 eV \cite{d2006optical} are considered. 
With these values, the unstrained system is predicted to become direct around 7\% Sn content. 
This value is consistent with those reported earlier \cite{polak2017electronic}. 
As the equilibrium solubility of Sn in Ge is below 1 at.\% \cite{thurmond1956germanium}, non-equilibrium and strain relaxation growth protocols are used to grow alloys at higher Sn content \cite{assali2019enhanced}. 

Controlling the strain in Ge$_{1-x}$Sn$_x$ alloys allows for further engineering of the band structure. Strain in the lattice leads to a hydrostatic shift of the conduction band critical points \cite{van1989band}:
\begin{align}
    \Delta E^\epsilon_i = 2 \epsilon_\| \alpha^i \left(1-\frac{2C_{12}}{C_{11}}\right),
\end{align}
where $\alpha^i$ is the hydrostatic deformation potential of the critical point, and $C_{11}$/$C_{12}$ are the elastic constants. In the case of strictly bi-axial strain perpendicular to the [001] growth direction, the eightfold $L$-valley degeneracy is not lifted. The degeneracy of the valance band at the $\Gamma$ point is first lifted by spin-orbit effects (SO separated from HH and LH), and further lifted by strain \cite{van1989band,terrazos2021theory}. Additionally, quantum confinement independently lifts the valance band degeneracy.

The band alignment and confinement in Ge$_{1-x}$Sn$_{x}$/Ge QWs is investigated over a parameter-space consisting of a Sn content in the 0-20 at.\% range and an in-plane compressive strain in the 0-2\% range. Here, the absolute values of strain are used.
By examining only the compressively strained QWs, we ensure that the top of the valance band in the QW has a HH character. Since the lattice parameter of Sn is larger than that of Ge ($a_{Sn}=6.48$ \AA  
 \cite{madelung2012semiconductors} , $a_{Ge}=5.65$ \AA
  \cite{reeber1996thermal}), this means that the Ge barriers are under a tensile strain, corresponding to a top valance band of LH character.
 
\begin{figure}
\includegraphics[width=\linewidth]{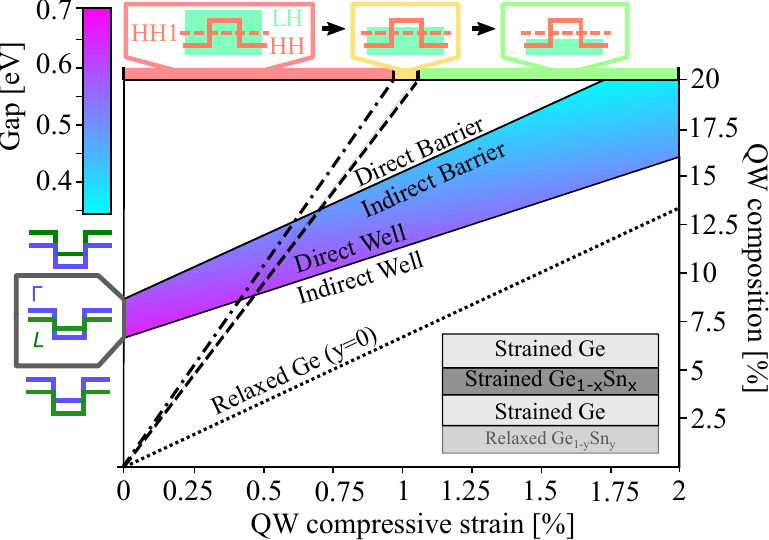}
\caption{\label{fig:QW_type_map} 
Map of the bandgap directness and hole confinement in the Ge$_{1-x}$Sn$_{x}$/Ge.
The solid lines represent the crossing between direct and indirect gap for the Ge barrier (top line) and the Ge$_{1-x}$Sn$_{x}$ well (bottom line), respectively. The bandgap energy is given in the region where the well is direct and the barriers are indirect. Left of the dot-dashed line, the HH band is bellow the LH continuum. Right of the dashed line, the first HH state is above the LH continuum for a 10 nm-thick well. On the dotted line, the system is reduced to a pseudomorphic Ge$_{1-x}$Sn$_x$ on relaxed Ge QW. Top insets show the evolution of the VB, while bottom-left insets show the evolution of the conduction bands.
}
\end{figure}

The results obtained are shown in Fig. \ref{fig:QW_type_map}, summarizing the electronic structure and the nature of confinement
in Ge$_{1-x}$Sn$_{x}$/Ge QWs in the strain-composition space.

The alignment of the conduction band is shown in the left inset, with green and blue for the $L$ and $\Gamma$ critical points, respectively. The region where the QW gap is direct and the barriers are indirect is colored as the bandgap energy color scale bar.  In the absence of strain, the Ge$_{1-x}$Sn$_{x}$ QW is directly around the 7 at.\% Sn content. The required Sn content increases with increasing strain, roughly following a  4.5 at.\% increase in Sn content for every 1\% increase in the absolute value of $\epsilon_\parallel$.

The alignment of the valence band is shown in the top inset of Fig. \ref{fig:QW_type_map}, with the HH and LH bands in orange and light green, respectively. The latter is mostly a continuum in the barriers due to the strain distribution.
At low strain, the HH band is buried under the LH continuum (red zone). To lift the HH band above the LH continuum (yellow zone), an increase of roughly 0.048$\%$ in $\epsilon_\parallel$ for each increase of 1at.\% in Sn content is required; as indicated by the dot-dashed line in Fig. \ref{fig:QW_type_map}. However, this does not guarantee that the first HH confined level will be above the continuum (green zone). A slightly higher strain is needed. For example, an increase of 0.053 $\%$ in $\epsilon_\parallel$ for a 1 at.$\%$ increase in Sn content is necessary to bring the first HH confined level above the LH continuum for a 10 nm-thick QW. This is indicated by the dashed line in Fig. \ref{fig:QW_type_map}.

We find that there is a wide range of combinations of strain and composition that ensure a direct Type I QW confining both HH and electrons at $\Gamma$, while having indirect barriers (color and green zone overlap). This region allows for the modulation of the QW direct gap energy from 0.35 to 0.65 eV. In particular, within the strain composition space considered, obtaining a direct gap is impossible for pseudomorphic Ge$_{1-x}$Sn$_x$ grown on a relaxed Ge system (dotted line in Fig. \ref{fig:QW_type_map}). Instead, pseudomorphic growth on strain-relaxed Ge$_{\text{1-y}}$Sn$_\text{y}$ ($y< x$) buffer is needed to achieve the structures shown in Fig. \ref{fig:QW_type_map}. 

\subsection{Effective parameters of the QW states}

\begin{figure}
    \centering
\includegraphics[width=\linewidth]{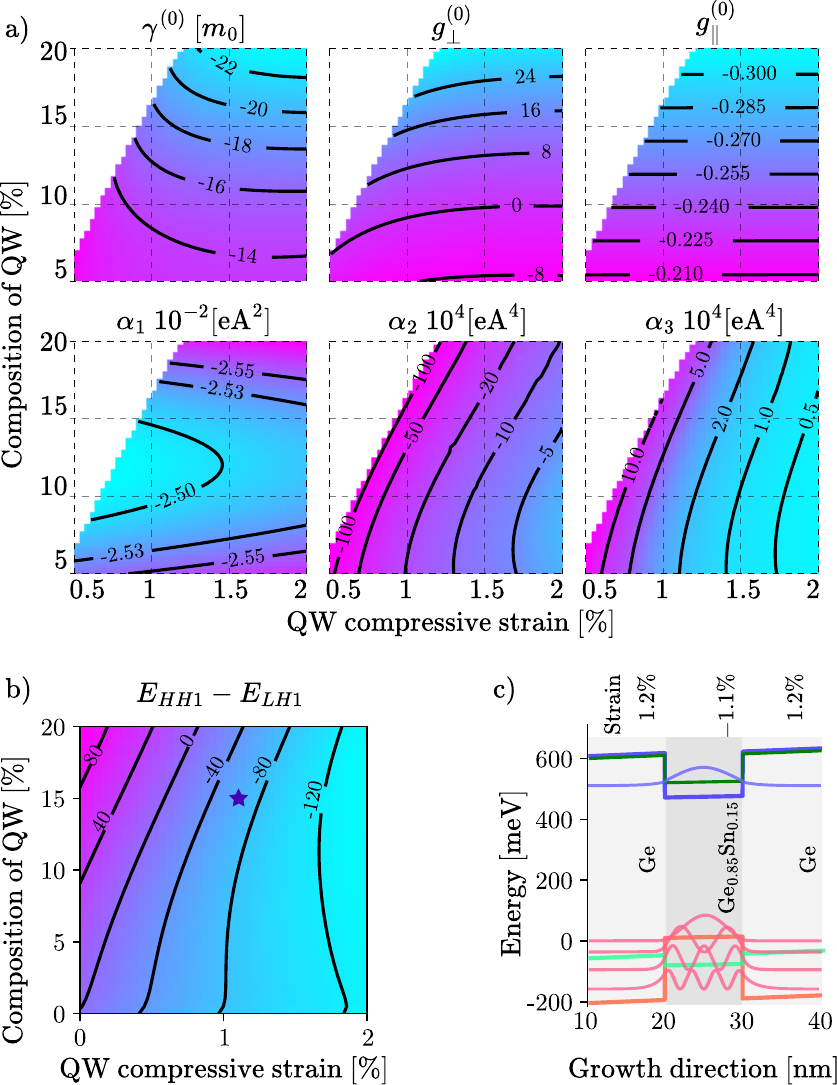}d
    \caption{a) Effective parameters for a 10 nm-thick strained Ge$_{1-x}$Sn$_x$/Ge QW
    b) Splitting between the first confined HH state and the LH continuum 
    c) Band alignment and confined states for a Ge$_{0.85}$Sn$_{0.15}$/Ge strained QW (purple star in b)) under 0.5 mV/nm electric field.}
    \label{fig:eff_params}
\end{figure}

The effective parameters for small $\mathbf{E}_z$ are evaluated by considering the constant terms for the effective mass and g-factor parameters,  
and the linear terms for the Rashba parameters:
\begin{align}
    &\gamma(E_z) = \gamma^{(0)}+O(E_z^2) \\
    &g^{\mathrm{QW}}_{\|/\perp}(E_z) = g_{\|/\perp}^{(0)}+O(E_z^2)\\
    &\beta_{1/2/3}(E_z) = \alpha_{1/2/3} E_z+ O(E_z^3) 
    \label{eq:beta(E)}
\end{align}
The Rashba parameters are a consequence of structural inversion asymmetry, they therefore vanish when no electric field is present in symmetrical QWs, as shown in Eq. (\ref{eq:beta(E)}). 
 Fig. \ref{fig:eff_params} (a) shows the effective parameters of the first confined HH state as a function of Sn content and strain in a 10 nm-thick QW with 20 nm-thick barriers, when considering 200 out-of-plane states. The graphs are left blank where the confined HH is buried under the LH continuum.
The splitting between the first confined HH state and the LH continuum is shown in Fig. \ref{fig:eff_params} (b). 

Due to the crossings between the HH subband and the LH continuum, an asymptotic behavior occurs for all effective parameters except $g_\parallel^{\mathrm{QW}}$ and $\beta_1$ when Sn content and strain are near the 0 splitting (also seen as the dashed line in Fig. \ref{fig:QW_type_map} ). $g^{(0)}_\|\propto \bra{h} q \ket{h }$ is not sensitive to strain as it depends only on the shape of the envelopes.
Both $g^{(0)}_\|$ and $\alpha_1$ are very small, as they are directly proportional to the anisotropic contribution to the g-factor $q$. The $q$ parameter is higher in Sn than in  Ge \cite{lawaetz1971valence}, thus slightly enhancing the in-plane g-factor by 0.075 for every  10 at.\% increase in Sn content. Because the $\kappa$ parameter of Ge and Sn are of different signs, $g^{(0)}_\perp\propto \bra{h}6 \kappa + 27 q/2 \ket{h}$ vanishes for certain concentration and strain values. 

As the split between the HH state and the LH continuum grows, $\alpha_2$ and $\alpha_3$ go to 0, as they are directly proportional to its inverse.
This splitting is directly related to compressive strain, which highlights once more the fact that Ge$_{1-x}$Sn$_x$/Ge QWs should be grown on a substrate that alleviates compressive strain in the QW to ensure a high SOC. 
The limiting effect of high compressive strain due to the high Sn content on the Rashba coupling has been observed experimentally \cite{tai2021strain}. 
Since $\alpha_3 \propto \bra{h} \gamma_2-\gamma_3 \ket{l}$ and $\alpha_2 \propto \bra{h} \gamma_2+\gamma_3 \ket{l} $, the latter is one order of magnitude higher. 
The Dresselhauss coupling terms are absent in group IV semiconductors due to the lack of bulk inversion symmetry. It should be noted that linear Rashba would be enhanced by strain gradients \cite{abadillo2023hole} due to gate electrode deposition, for example, and atomic-level interface details \cite{liu2022emergent}, both of which are not considered here.

\subsection{g-factor of the QD}
In the following, the out-of-plane confinement is achieved in a 10 nm-thick Ge$_{0.85}$Sn$_{0.15}$ QW with 20 nm-thick Ge barriers. The strain is set at -1.1\% in the QW layer, which corresponds to a tensile strain of 1.2\% in the barriers. This can be achieved by growing pseudomorphically on a Ge$_{0.92}$Sn$_{0.08}$ strain-relaxed buffer layer. This system, shown in Fig. \ref{fig:eff_params} (c), was selected because it provides a direct band gap QW, confinement of HH and electrons at $\Gamma$, a high $\alpha_2$, and a non-zero $g_\perp^{(0)}$. The out-of-plane electric field is kept constant at 0.5 mV/nm. We consider 64 in-plane ($n_{x/y}\leq 8 $) and 200 out-of-plane states.
The g-factor of the QD is:
\begin{equation}
g^{\mathrm{QD}}=\frac{\Delta E}{\mu_B B},    
\end{equation}
where $\Delta E = E_\mathbb{1}-E_\mathbb{0}$ is the difference in energy between the first $| \mathbb{0}\rangle$ and second  $|\mathbb{1}\rangle$ states of the QD whose Hamiltonian is given by Eq. (\ref{eq:HamiltonianQD}).

\begin{figure}
    \centering
    \includegraphics[width=0.95\linewidth]{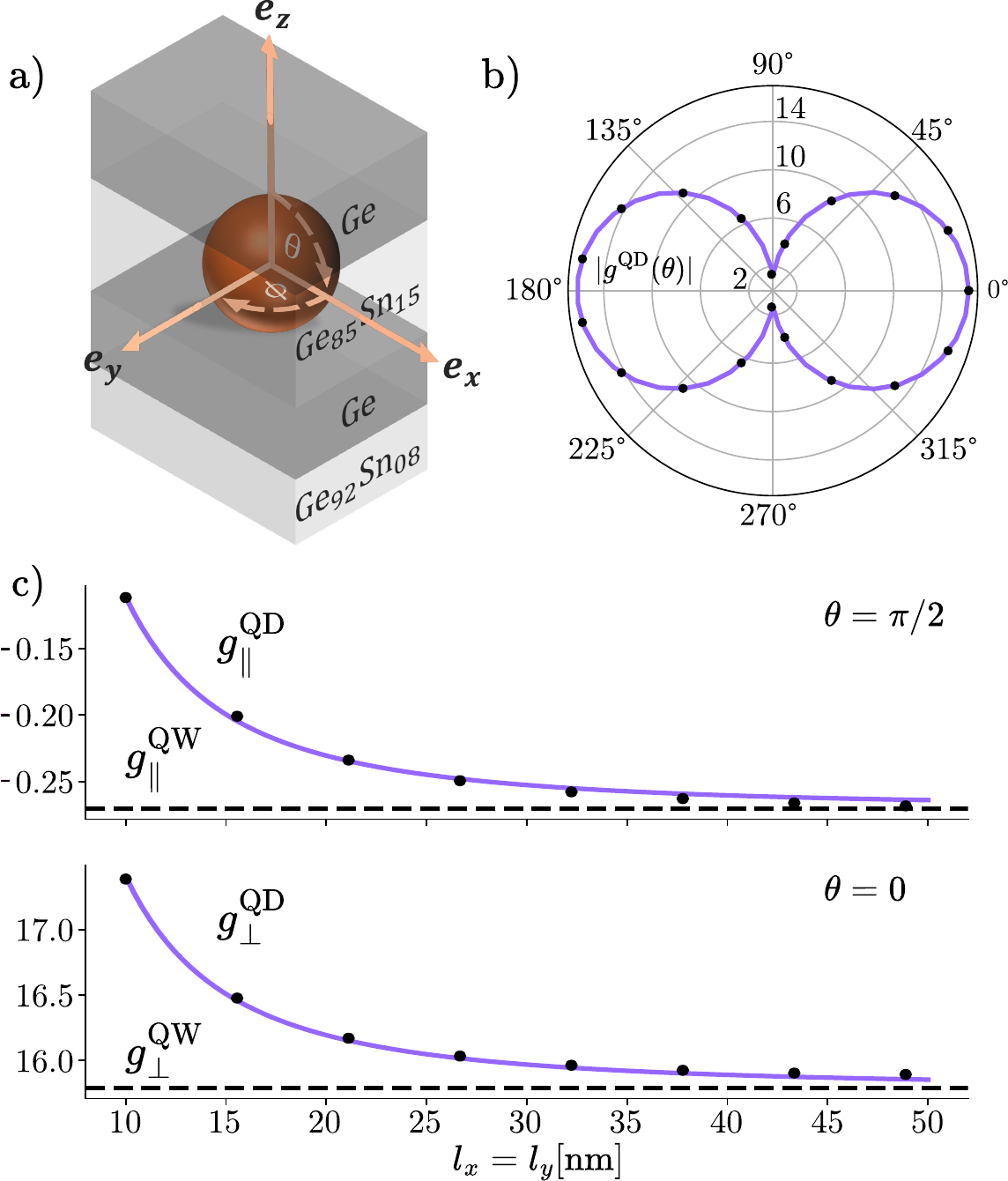}
    \caption{ a) Sketch of the heterostructure with the axis system showing a confined HH QD b) Polar diagram for the g-factor of a symmetric 25 nm QD  c) in-plane and out-of-plane g-factor as a factor of QD size. Dots in b) and c) represent the results of the diagonalisation of the full 3D Hamiltonian, and blue lines are Eq. (\ref{eq:gQDtheta}) for panel b) and Eq. (\ref{eq:gQDr0}) for panel c). }
    \label{fig:gfactor}
\end{figure}

The g-factor of the QD for both in-plane and out-of-plane magnetic field configurations is shown in Fig. \ref{fig:gfactor}. The in-plane g-factor is 2 orders of magnitude lower than that of the out-of-plane. For a large QD radius, both factors tend asymptotically to the QW values. To explain the relationship between $g^{\mathrm{QD}}$ and $g^{\mathrm{QW}}$, the third-order SWT is not sufficient. An extra coupling term, which would only be fully captured by a fourth-order SWT \cite{DelVecchio_2023}, must be added:

\begin{equation}
    \mathbf{H}_{\chi} =\frac{\alpha _{0}}{2l_{B}^{2}} K_{\parallel }^{2} \boldsymbol{\sigma}\cdot \boldsymbol{\chi} \cdot \mathbf{n},  
\end{equation}
where the $\boldsymbol{\chi}$-matrix is:

\begin{equation}
\mathbf{\chi}=\begin{bmatrix}
\chi^{\mathrm{QW}}_{||} & 0 & 0\\
0 & \chi^{\mathrm{QW}}_{||} & 0\\
0 & 0 & \chi^{\mathrm{QW}}_{\perp }
\end{bmatrix}
\end{equation}

Treating the extra term $\mathbf{H}_{\mathrm{\chi}}$ as a perturbation of $\mathbf{H}_{\mathrm{FD}}$ to the first order yields:

\begin{equation}
\label{eq:gQDr0}
    g^{\mathrm{QD}}_{\|/\perp} \approx g^{\mathrm{QW}}_{\|/\perp}+\frac{\chi^{\mathrm{QW}}_{\|/\perp}}{l^2},
\end{equation}
where it can now be seen that $\mathbf{H}_{\mathrm{\chi}}$ acts as a momentum-dependent correction to the g-factor \cite{bosco2021squeezed}, explaining the relationship between the size of the QD and its g-factor.  As a function of the magnetic field angle, the QD g-factor is easily found by solving $\mathbf{H}_{\mathrm{FD}}$:

\begin{align}
\label{eq:gQDtheta}
    |g^{\mathrm{QD}}(\theta)| = \sqrt{\left(g^{\mathrm{QD}}_\perp \cos \theta \right)^2+\left(g^{\mathrm{QD}}_\|\sin \theta\right)^2}
\end{align}

\subsection{Dipole moment}

At the resonant condition $\Delta E = h \nu$, an in-plane alternating electric field $\mathbf{E}_{\text{ac}}=E_{\text{ac}}\mathbf{e}_n$, where  $\mathbf{e}_n = \cos \varphi \mathbf{e}_x+ \sin \varphi \mathbf{e}_y$ is the unit vector parallel to $\mathbf{E}_{\text{ac}}$,  generates a Rabi frequency:
\begin{align}
     \Omega=E_{a c} \frac{d}{\hbar},
\end{align}
where $d=e|\langle \mathbb{0}|\boldsymbol{r}\cdot \mathbf{e}_n| \mathbb{1}\rangle|$ is the dipole moment.
It is calculated by using the ladder operators given by Eq. (\ref{eq:ladder}) to write $\boldsymbol{r}$:

\begin{equation}
\begin{split}
   \frac{d}{e}= |\bra{\mathbb{0}}\frac{l_x \cos \varphi}{\sqrt{2}}\left(a_x+a_x^\dagger\right) +\frac{l_y \sin \varphi }{\sqrt{2}} \left(a_y+a_y^\dagger\right)\ket{\mathbb{1}}|
   \end{split}
\end{equation}

From this equation, we can see that the electric field will only couple levels that have $\Delta n_{x,y}=\pm 1$, since $\boldsymbol{r}$ is linear in $a_{x/y}$. Consequently, treating the Rashba term $\mathbf{H}_{\mathrm{R}}$ as a perturbation of the effective $\mathbf{H}_{\text{FD}}$ reveals that $d$ is directly proportional to $\beta_3\propto K_+K_-K_+$, but does not depend on $\beta_2 \propto K_+^3$ \cite{terrazos2021theory,wang2021optimal}. Since it is very weak, we can neglect $\beta_1 \propto K_-$ for this system. As a result of this perturbation, we obtain:

\begin{equation}
    d^{\mathrm{eff}} \approx \tilde{d}^{\mathrm{eff}} = 2 A \frac{\beta_3 g^{QD} \mu_B B \tilde{m}^2 l^2}{\hbar^4},
\end{equation}
where $d^{\mathrm{eff}}=e|\langle \mathbb{0}^{\mathrm{eff}}|\boldsymbol{r}\cdot \mathbf{e}_n| \mathbb{1}^{\mathrm{eff}}\rangle|$, $| \mathbb{0}^{\mathrm{eff}}\rangle$ and $ | \mathbb{1}^{\mathrm{eff}}\rangle $ are the first and second states of (\ref{eq:effQD}), and $0 \leq A \leq 1$ is a constant that depends on the angle between the in-plane electric field and the magnetic field. In particular, $A$ is always 1 for $\theta = 0$. When $\theta=\pi/2$, $A$ will be 1 if $\mathbf{B}\parallel \mathbf{E}_\mathrm{ac}$ and 0 when $\mathbf{B}\perp \mathbf{E}_\mathrm{ac}$.

\begin{figure}
\centering
\includegraphics[width=\linewidth]{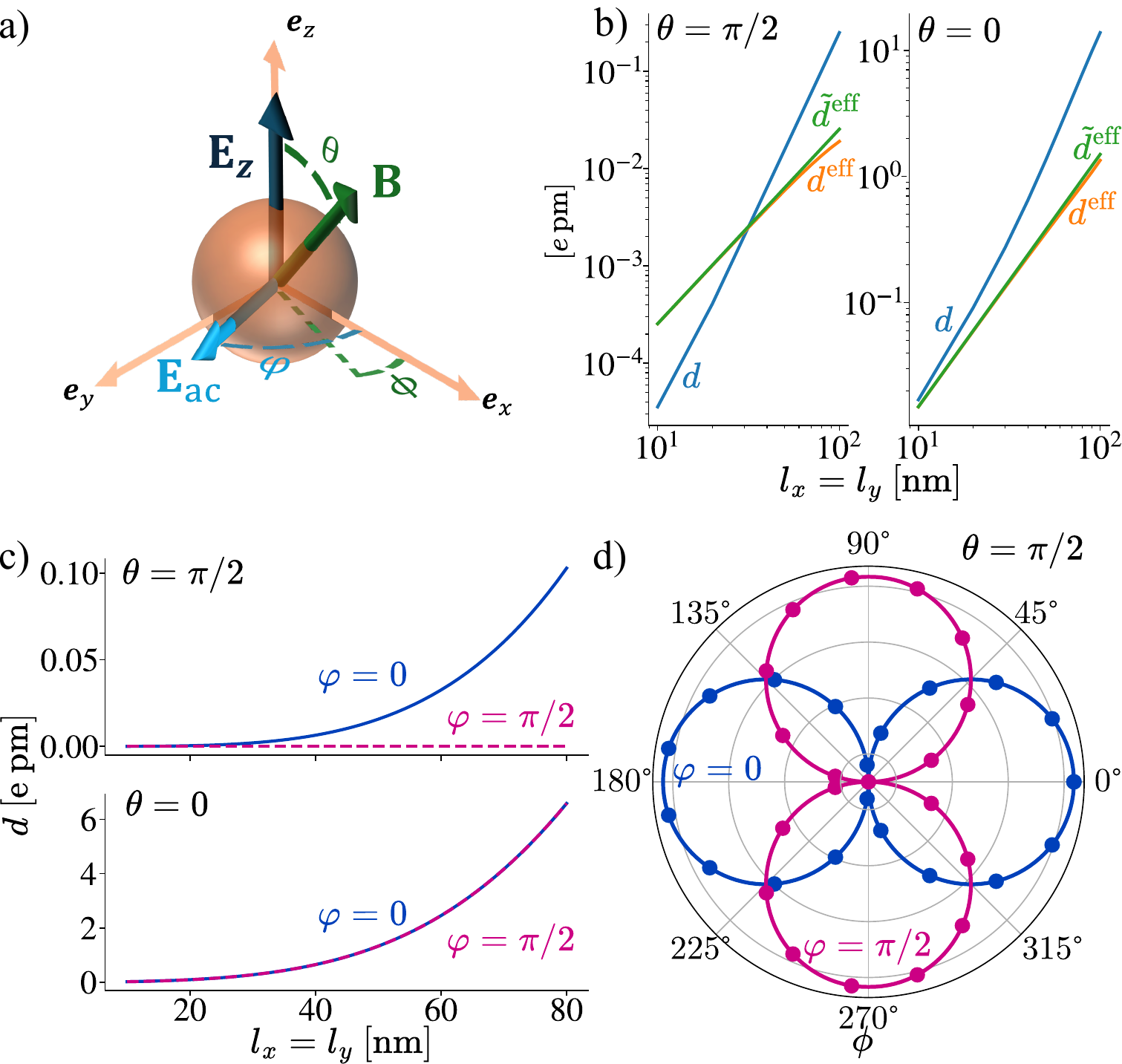}
     \caption{Dipole moment of a QD inside a 10 nm strained Ge$_{0.85}$Sn$_{0.15}$ QW under $B$=0.06 T, $E_z$=0.5 mV/nm, and $E_{\mathrm{ac}}\propto \zeta = \gamma$. a) Sketch of the QD with the relevant fields applied. b) Comparison between the solution of the full Hamiltonian $d$, of the effective Hamiltonian $d^{\mathrm{eff}}$, and the latter's perturbative solution $\tilde{d}^{\mathrm{eff}}$. c) $d$ for in-plane and out-of-plane magnetic field configurations when $\phi=0$. d) $d$ for different in-plane field orientations for a 25 nm QD, where each graduation represent 0.25 e$\cdot$Fm }
    \label{fig:dso}
\end{figure}

We first compare the dipole moment from the full Hamiltonian, the effective Hamiltonian, and the perturbative approximation of the latter, shown in Fig. \ref{fig:dso} b). We find that $\tilde{d}^{\mathrm{eff}}$ is a very accurate approximation of $d^\mathrm{eff}$. In contrast, $d^{\mathrm{eff}}$ does not seem to accurately describe the dipole moment computed from the full Hamiltonian.
For the out-of-plane configuration, it is accurate for small $r$, but we find that eventually $d\propto r^4$, a behavior that is not captured by the effective Hamiltonian. In the in-plane configuration, we find that the $r^4$ behavior occurs at lower $r$, making the effective Hamiltonian a worse approximation. 

In both cases, the effective Hamiltonian does produce the good order of magnitude for $d$, allowing us to understand certain trends.  
Notably, it allows one to attribute the two orders of magnitude difference between the in-plane and out-of-plane $d$, shown in Fig. \ref{fig:dso} c), to the fact that $\tilde{d}^{\mathrm{eff}}\propto g^{\mathrm{QD}}$.
Fig \ref{fig:dso} d) gives the result of $d$ for various in-plane orientations. Under these orientations, the dipole moment vanishes when the magnetic field is perpendicular to the driving field \cite{sarkar2023electrical}, and it is maximal when they are aligned. We find that $d$ is expected to be on the order of 0.01 e pm and 1 e pm for the in-plane and out-of-plane configurations, respectively.

\subsection{Relaxation rate}
We developed a model to study the behavior of phonon-assisted relaxation rate in gate-defined Ge$_{1-x}$Sn$_{x}$ QDs.
By first separating the out-of-plane and in-plane part of the hole-phonon interaction Hamiltonian, the relaxation rate derived from Fermi's golden rule is \cite{li_2020} (see Appendix \ref{sec:relaxationDerivation}):
\begin{equation}
\begin{split}
        \Gamma = R( \omega,T) \sum_\alpha \frac{1}{v_\alpha^5} \int d\theta d\varphi \sin \theta  |\langle \mathbb{0}| \mathrm{e}^{i q_\alpha \boldsymbol{r}} \mathbf{W}_z (\epsilon_{\alpha }) |\mathbb{1}\rangle |^2,
        \label{eq:rate}
    \end{split}
\end{equation}
where $\hbar \omega = \Delta E$, $R$ is a prefactor, $\alpha=\{l,t_1,t_2\}$ is the branch index, $T$ is the temperature, $v_\alpha$ is the speed of sound, $q_\alpha=\omega/v_\alpha$, $\mathbf{W}_z$ is the out-of-plane interaction Hamiltonian, and $\epsilon$ is the phonon deformation for the different branches, given in Appendix \ref{sec:relaxationDerivation}. To solve Eq. (\ref{eq:rate}), the exponential is expanded as:

\begin{equation}
    \mathrm{e}^{i q_\alpha \boldsymbol{r}} \mathbf{W}_z (\epsilon_{\alpha \boldsymbol{q}}) \approx (1+i q_\alpha \boldsymbol{r} - \frac{q_\alpha^2}{2} \boldsymbol{r}^2 ) \mathbf{W}_z (\epsilon_{\alpha \boldsymbol{q}})
    \label{eq:expension}
\end{equation}

\begin{figure}
    \centering
    \includegraphics[width=\linewidth]{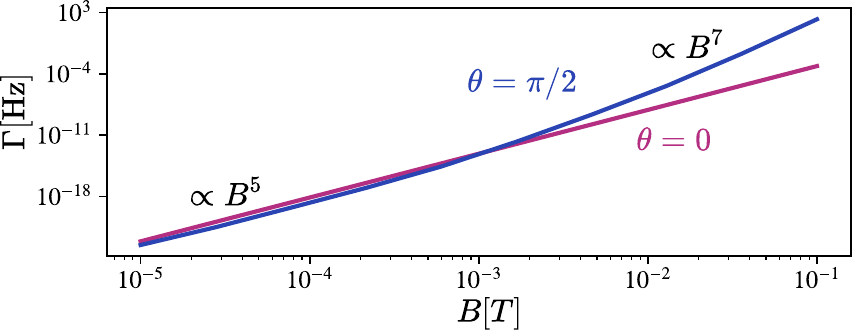}
    \caption{Phonon-assisted relaxation rate for both in-plane and out-of-plane magnetic field configurations for a 25 nm QD}
    \label{fig:relax}
\end{figure}

Fig. \ref{fig:relax} exhibits the obtained behavior of the relaxation rate for both in-plane and out-of-plane configurations.
Since  $R(T=0) \propto \omega^3$, the relaxation rate is at least $\propto B^3$. Both $|\mathbb{0}\rangle$ and $|\mathbb{1}\rangle $ can be expanded as a series of eigenstates of $\mathbf{H}_{\text{QD}}(B=0)$, which is dominated by constant in B term and an admixture of linear-in-B terms:
\begin{align}
 &|\mathbb{0}\rangle = |\uparrow 0\rangle + B \sum_{\sigma n} C^{\mathbb{0}}_{\sigma,n} |\sigma n \rangle  + ... \\ 
 &|\mathbb{1}\rangle = |\downarrow 0\rangle + B \sum_{\sigma n} C^{\mathbb{1}}_{\sigma,n} |\sigma n \rangle  + ...
\end{align} 
 The $B^5$ term arises from the first term of the expansion (\ref{eq:expension}) and the fact that $|\langle \mathbb{0} | \mathbf{W}_z | \mathbb{1} \rangle|^2$ vanishes when the states are time-reversal symmetric, meaning $\langle \uparrow0 | \mathbf{W}_z |\downarrow 0\rangle \approx 0$. The $B^7$ behavior arises from the second term of the expansion that also includes an extra $q_\alpha^2 \propto B^2$. The latter is negligible in the in-plane configuration. At a finite temperature, the rate for low magnetic field will depart from the $B^5$ behavior due to the prefactor \cite{li_2020}. 

\section{Conclusion}
In summary, this work leverages strain- and composition-dependent band structure of Ge$_{1-x}$Sn$_{x}$ semiconductors to introduce silicon-compatible hole qubits. These  Ge$_{1-x}$Sn$_{x}$ qubits exhibit all the characteristics that make Ge an attractive system while providing a tunable direct bandgap across a broad range of energies. The effective properties of HH confined in strained Ge$_{1-x}$Sn$_{x}$/Ge quantum wells were studied, indicating that a high compressive strain is detrimental as it lowers Rashba coupling and keeps the bandgap indirect. We propose growing on relaxed Ge$_{1-y}$Sn$_{y}$ ($x>y$) to circumvent this limitation. Moreover, we have demonstrated that this material system can be used as a platform for EDSR in electrostatically defined quantum dots, and the key parameters shaping their performance were discussed. To this end, we have developed and compared an effective approach with the diagonalization of the full Hamiltonian, allowing insights into the behavior of the dipole moment, $d$. We expect $d$ to vary between 0.01 and 1 e pm for in-plane and out-of-plane magnetic field configurations, respectively. 
A model for phonon assisted relaxation rate, $\Gamma$ , in gate-defined quantum dots was implemented, showing that $\Gamma$ should be $\propto B^5$ for both field configurations, and become $\propto B^7$ under the out-of-plane field beyond a  threshold of about $B$=0.002 T. Finally, it is important to note that, unlike Ge low-dimensional systems, alloy scattering in Ge$_{1-x}$Sn$_{x}$ is expected to reduce carrier mobilities. However, recent studies  \cite{tai2021strain} indicate that hole mobility in Ge$_{1-x}$Sn$_{x}$ surpasses the threshold for implementing hole spin qubits  \cite{watzinger2018germanium}.

\medskip
\noindent {\textbf{Acknowledgments}}.
O.M. acknowledges support from NSERC Canada (Discovery, SPG, and CRD Grants), Canada Research Chairs, Canada Foundation for Innovation, Mitacs, PRIMA Qu\'ebec, Defence Canada (Innovation for Defence Excellence and Security, IDEaS), the European Union’s Horizon Europe research and innovation program under Grant Agreement No 101070700 (MIRAQLS), the US Army Research Office Grant No. W911NF-22-1-0277, and the Air Force Office of Scientific and Research Grant No. FA9550-23-1-0763. \\

\appendix
\section{Kane Hamiltonian}
\label{sec:kaneHamiltonian}
For simplicity, the elements of the Kane Hamiltonian are given in the cartesian representation $\{|S\uparrow\rangle,|S\downarrow\rangle,|X\uparrow\rangle,|Y\uparrow\rangle,|Z\uparrow\rangle,|X\downarrow\rangle,|Y\downarrow\rangle,|Z\downarrow\rangle\}$. The first term $\mathbf{H}_k$  is given by:
\begin{equation}
      \mathbf{H}_k = \begin{bmatrix}
                    \mathbf{1}_{2}\otimes\mathbf{H}^k_{cc} & \mathbf{1}_{2}\otimes\mathbf{H}^k_{cv}  \\                    \mathbf{1}_{2}\otimes\mathbf{H}_{cv}^{k\dagger}  & \mathbf{1}_{2}\otimes\mathbf{H}^k_{vv}  
    \end{bmatrix}
\end{equation}
\begin{align}
  &\mathbf{H}^k_{cv}=\begin{bmatrix}
       iP K_x & iP K_y&  iP K_z 
  \end{bmatrix} \\
  & \mathbf{H}^k_{cc} = E_g + \sum_{\alpha} K_{\alpha}^\dagger A K_\alpha + \alpha_0 \sum_{\alpha \beta \gamma} \epsilon_{\alpha\beta\gamma}\sigma_\alpha K_\beta^\dagger (g-g_0) K_\gamma \\
  & (\mathbf{H}^k_{vv})_{\alpha,\beta} = \begin{cases}
       \sum_{\gamma} K_\gamma^\dagger M K_\gamma + K_\alpha^\dagger(L-M) K_\alpha \quad  & \alpha = \beta \\
       K_\alpha^\dagger N_+ K_\beta + K_\beta N_- K_\alpha \quad & \alpha \neq \beta
  \end{cases}
\end{align}
where $\epsilon_{\alpha\beta\gamma}$ is the Levi-Civita symbol, $E_g$ is the bandgap, $g_0 \approx 2$ the g-factor of the free electron,  $P$ the Kane momentum matrix element, $A$ and $g$ are the conduction band parameters,  $[\alpha,\beta,\gamma]$ all span the cartesian directions $\{X,Y,Z\}$ and the parameters [$L$,$M$,$N_+$,$N_-$] are related to the rescaled Luttinger parameters $\gamma_1$ $\gamma_2$, $\gamma_3$ and to $\kappa$ by:
\begin{equation}
    \begin{bmatrix}
        L \\ M \\ N_+ +\frac{\hbar^2}{2 m_0} \\ N_- -\frac{\hbar^2}{2 m_0}
    \end{bmatrix} = -\frac{\hbar^2}{2 m_0} \begin{bmatrix}
        1 & 4 & 0 & 0 \\
        1 &-2 & 0 & 0 \\
        0 & 0 & 3 & 3 \\
        0 & 0 & 3 &-3 
    \end{bmatrix}
    \begin{bmatrix}
        \gamma_1 \\ \gamma_2 \\ \gamma_3 \\ \kappa
    \end{bmatrix}
\end{equation}
The rescaling relations that relates $\gamma_{1,2,3},\kappa,A,g$ to the parameters of the 6-band Luttinger-Kohn Hamiltonian $\gamma_{1,2,3}^L,\kappa^L$ and to the 2-band parabolic conduction band Hamiltonian $m^*,g^*$ are \cite{Winkler2003}:
\begin{align}
      \gamma_1 = \gamma_1^L - \frac{1}{3} \frac{P^2}{ \alpha_0 E_g} \label{eq:gamma1L}\\
      \gamma_{2,3} = \gamma_{2,3}^L - \frac{1}{6}\frac{P^2}{ \alpha_0 E_g} \label{eq:gamma23L}\\
      \kappa = \kappa^L - \frac{1}{6} \frac{P^2}{\alpha_0 E_g} \label{eq:kappaL}
\end{align}
\begin{align}
        g = g^* + \frac{2}{3} \frac{P^2}{ \alpha_0 E_g} 
    -\frac{1}{3}\frac{P^2}{\alpha_0 (E_g+\Delta)}\label{eq:geff}\\
    A = \frac{m_0}{ \alpha_0 m^*} - \frac{2}{3}\frac{P^2}{E_g}-\frac{1}{3}\frac{P^2}{E_g+\Delta} \label{eq:meff}
\end{align}
where $\Delta$ is the split-off gap. The SOC matrix $\mathbf{H}_{SO}$ is given by:
\begin{align}
          &\mathbf{H}_{SO} = \frac{\Delta}{3}\begin{bmatrix}
      \mathbf{1}_2 & \mathbf{0} \\
      \mathbf{0}   & \mathbf{H}^{SO}_{vv}
      \end{bmatrix} \\
        &\mathbf{H}^{SO}_{vv} = \begin{bmatrix}
        0 & -i & 0 & 0 & 0 & 1 \\
        i & 0  & 0 & 0 & 0 &-i \\
        0 & 0  & 0 &-1 & i & 0 \\
        0 & 0  &-1 & 0 & i & 0 \\
        0 & 0  &-i &-i & 0 & 0 \\
        1 & i  & 0 & 0 & 0 & 0 
    \end{bmatrix}
\end{align}
The coupling to the magnetic field $\mathbf{H}_B$ is given by:
\begin{align}
    &\mathbf{H}_B = \begin{bmatrix}
          \mathbf{H}_B^k & \mathbf{0} \\ 
          \mathbf{0}     & \mathbf{H}_B^k \otimes \mathbf{1}_3
      \end{bmatrix}\\
         &\mathbf{H}_B^k  = \mu_B \frac{g_0}{2} \left(\sigma_x B_x + \sigma_y B_y + \sigma_z B_z \right)
\end{align}
The strain matrix $\mathbf{H}_\varepsilon$ \cite{Bir_1974} is:
\begin{align}
    &\mathbf{H}_\varepsilon = \begin{bmatrix}
          \mathbf{1}_2 \otimes \mathbf{H}^\varepsilon_{cc} & \mathbf{0} \\
          \mathbf{0} & \mathbf{1}_2 \otimes \mathbf{H}^\varepsilon_{vv}
      \end{bmatrix}
      \label{eq:kpStrain} \\
    &\mathbf{H}_{cc}^\varepsilon = a_c \mathrm{Tr}\{\stackrel{\leftrightarrow}{\varepsilon}\} \\
    &(\mathbf{H}_{vv}^\varepsilon)_{\alpha \beta} = \begin{cases}
        m \mathrm{Tr}\{\stackrel{\leftrightarrow}{\varepsilon}\}+(l-m)\varepsilon_{\alpha\alpha} & \alpha = \beta \\
        n \varepsilon_{\alpha\beta} & \alpha \neq \beta
    \end{cases}
\end{align}
where $a_c$ is the conduction band deformation potential and the 
valance band deformation potentials $l$,$m$ and $n$ are related to the more common parameters $a_v$, $b$ and $d$ by:
\begin{equation}
    \begin{bmatrix}
        l \\ m \\ n 
    \end{bmatrix} =
    \begin{bmatrix}
        1 & 2 & 0 \\
        1 &-1 & 0 \\
        0 & 0 &\sqrt{3}
    \end{bmatrix}
    \begin{bmatrix}
        a_v \\  b \\ d 
    \end{bmatrix}
\end{equation}
The Hamiltonian in the angular momentum representation 
$\{|CB\uparrow\rangle,|CB\downarrow\rangle,|HH\uparrow\rangle,|LH\uparrow\rangle,|LH\downarrow\rangle,|HH\downarrow\rangle,|SO\uparrow\rangle,|SO\downarrow\rangle\}$ is obtained from the transformation  \cite{sun2009strain}:
\begin{align}
& \mathbf{H}_{\text{QW}} = \boldsymbol{U}^\dagger \mathbf{H}^{xyz}_{\text{QW}} \boldsymbol{U}\\
     &\boldsymbol{U} = \begin{bmatrix}
              \boldsymbol{1}_2 & \boldsymbol{0} \\
              \boldsymbol{0}   & \boldsymbol{U}_{vv}
\end{bmatrix}
\end{align}
\begin{align}
          &\boldsymbol{U}_{vv} = \begin{bmatrix}
        \frac{-1}{\sqrt{2}} & 0 & \frac{1}{\sqrt{6}} & 0 & 0 & \frac{-1}{\sqrt{6}} \\
        \frac{-i}{\sqrt{2}} & 0 & \frac{-i}{\sqrt{6}}& 0 & 0 & \frac{i}{\sqrt{3}} \\
        0 & \sqrt{\frac{2}{3}} & 0 & 0 & \frac{-1}{\sqrt{3}} & 0 \\
        0 & \frac{-1}{\sqrt{6}}& 0 & \frac{1}{\sqrt{2}} & \frac{-1}{\sqrt{3}} & 0 \\
        0 & \frac{-i}{\sqrt{6}}& 0 &\frac{-i}{\sqrt{2}} & \frac{-i}{\sqrt{3}} & 0 \\
        0 & 0   & \sqrt{\frac{2}{3}} & 0 & 0 & \frac{1}{\sqrt{3}}
    \end{bmatrix}
\end{align}
The cubic in $\mathbf{J}$ correction to the g-factor in the momentum basis is given by:

\begin{align}
    &\mathbf{H}_q = \begin{bmatrix}
              \boldsymbol{0} & \boldsymbol{0} & \boldsymbol{0} \\
              \boldsymbol{0}   & -2 i \alpha_0 \sum_{\alpha \beta \gamma} (J_{\alpha })^3 K_\beta  \ q \  K_\gamma \varepsilon_{\alpha \beta \gamma } &\boldsymbol{0} \\
              \boldsymbol{0} &\boldsymbol{0} &\boldsymbol{0}
          \end{bmatrix} 
\end{align}
\begin{align}
    &J_x =\frac{1}{2} \begin{bmatrix}
        0 & \sqrt{3} & 0 & 0 \\
        \sqrt{3} & 0 & 2 & 0 \\
        0 & 2 & 0 & \sqrt{3} \\
        0 & 0 & \sqrt{3} & 0 
    \end{bmatrix}
\end{align}
\begin{align}
    &J_y = \frac{i}{2}\begin{bmatrix}
        0 & -\sqrt{3} & 0 & 0 \\
        \sqrt{3} & 0 & -2 & 0 \\
        0 & 2 & 0 & -\sqrt{3} \\
        0 & 0 & \sqrt{3} & 0
    \end{bmatrix}
\end{align}
\begin{align}
    &J_z = \frac{1}{2} \begin{bmatrix}
        3 & 0 & 0 & 0 \\
        0 & 1 & 0 & 0 \\
        0 & 0 & -1 & 0 \\
        0 & 0 & 0 & -3
    \end{bmatrix}
\end{align}

\section{Matrix in subband edge basis}
\label{sec:subbandedgebasis}

The matrix in the subband edge basis is obtained by projecting the QW Hamiltonian defined by Eq. (\ref{eq:QWHamiltonian}) on the enveloppes given by Eq. (\ref{eq:subband_enveloppes}). As the involved Hamiltonian depend on momentum operators that include the magnetic field, the final result is dependant on the choice of gauge. We give here the result when using the general vector gauge displayed in Eq. (\ref{eq:gauge3D}). We first define the following:
\begin{align}
&u_{\pm } =\{\gamma _{3} ,k_{z}\} \pm [ \kappa ,k_{z}]\\
&|\pm \rangle =|l\rangle \pm (\sqrt{2})^{\pm 1}|s \rangle \\
\end{align}
With bra-ket products representing integrals along the growth direction of position-dependant material parameters with envelopes functions, we define the following parameters:
\begin{widetext}

\begin{align}
&(\mathbf{g}_\perp^{\text{H}})_{HH} &&=-\bra{h} 6\kappa +\frac{27q}{2}\ket{h}\\
&(\mathbf{g}_\perp^{\eta })_{\eta \eta} &&=\bra{c} g\ket{c} -2\bra{+} \kappa \ket{+} -\frac{1}{2}\bra{\ell } q\ket{\ell } -\frac{4}{3}\left(\langle+|+\rangle -\langle-|-\rangle\right)\\
&(\mathbf{g}_{\parallel }^{\text{H}})_{HH} &&=-3\bra{h} q\ket{h}\\
&(\mathbf{g}_{\parallel }^{\eta })_{\eta\eta} &&=\bra{c} g\ket{c} -2\bra{+} \kappa \ket{-} -2\bra{-} \kappa \ket{+} -10\bra{\ell } q\ket{\ell } -\sqrt{2}\left(\langle-|s\rangle +\langle s|-\rangle\right)\\
&(\mathbf{g}_{\parallel }^{X})_{H\eta} &&=-2\sqrt{3}\bra{h}\left\{\kappa \ket{-} +\frac{7}{4} q\ket{\ell } -\frac{1}{\sqrt{2}}\ket{s}\right\}\\
&(\boldsymbol{\gamma }^{\text{H}})_{HH} &&=-\bra{h} \gamma _{1} +\gamma _{2}\ket{h}\\
&(\boldsymbol{\gamma }^{\eta })_{\eta\eta} &&=\bra{c}\frac{A}{\alpha _{0}}\ket{c} -\frac{1}{3}\bra{+} \gamma _{1} +\gamma _{2}\ket{+} -\frac{2}{3}\bra{-} \gamma _{1} -2\gamma _{2}\ket{-}\\
&(\mathbf{S})_{HH} &&=\mathbf{-}\frac{3i}{2}\bra{h}[ q,k_{z}]\ket{h}\\
&(\mathbf{R})_{\eta\eta} &&=\frac{1}{\sqrt{6} \alpha _{0}}\left(\bra{c} P\ket{+} +\bra{+} P\ket{c}\right) +i\left(\frac{1}{2}\bra{c}[ g,k_{z}]\ket{c} -\bra{+} u_{+}\ket{-} +\bra{-} u_{-}\ket{+} -5\bra{\ell }[ q,k_{z}]\ket{\ell }\right)\\
&(\mathbf{P})_{H\eta} &&=\bra{h}\left[\frac{P}{\sqrt{2} \alpha _{0}}\ket{c} -\sqrt{3} i\left( u_{+}\ket{-} +\frac{7}{4}[ q,k_{z}]\ket{\ell }\right)\right]\\
&(\boldsymbol{\mu })_{H\eta} &&=\sqrt{3}\bra{h} \frac{\gamma _{2} +\gamma _{3}}{2} \ket{+}\\
&(\boldsymbol{\delta })_{H\eta} &&=\sqrt{3}\bra{h}\frac{\gamma _{2} -\gamma _{3}}{2} \ket{+}\\
&(\mathbf{h}_{0})_{HH} &&=\mathbf{-}\bra{h} \gamma _{1} -2\gamma _{2}\ket{h}\\
&(\mathbf{h}_{1})_{HH} &&=\mathbf{-}\bra{h}\{\gamma _{1} -2\gamma _{2} ,k_{z}\}\ket{h}\\
&(\boldsymbol{\eta }_{0})_{\eta\eta} &&=\bra{c}\frac{A}{\alpha _{0}}\ket{c} -\frac{1}{3}\bra{+} \gamma _{1} -2\gamma _{2}\ket{+} -\frac{2}{3}\bra{-} \gamma _{1} +4\gamma _{2}\ket{-}\\
&(\boldsymbol{\eta }_{1})_{\eta\eta} &&=\frac{i}{\alpha _{0}}\sqrt{\frac{2}{3}}\left(\bra{c} P\ket{-} -\bra{-} P\ket{c}\right) +\bra{c}\left\{\frac{A}{\alpha _{0}} ,k_{z}\right\}\ket{c} -\frac{1}{3}\bra{+}\{\gamma _{1} -2\gamma _{2} ,k_{z}\}\ket{+} -\frac{2}{3}\bra{-}\{\gamma _{1} +4\gamma _{2} ,k_{z}\}\ket{-} \\
&(\mathbf{r})_{\eta\eta} &&=i\left(\bra{+} \gamma _{3}\ket{-} -\bra{-} \gamma _{3}\ket{+}\right)\\
&(\mathbf{p})_{H\eta} &&=i\sqrt{3}\bra{h} \gamma _{3}\ket{-},
\end{align}
where the indices $H$ and $\eta$ refer to the subband index. We define the following matrices:
\begin{align}
    \mathbf{E}_{0} =\begin{bmatrix}
\mathbf{E}^{\text{H}} & 0 & 0 & 0\\
0 & \mathbf{E}^{\eta } & 0 & 0\\
0 & 0 & \mathbf{E}^{\eta } & 0\\
0 & 0 & 0 & \mathbf{E}^{\text{H}}
\end{bmatrix} 
\end{align}
\begin{align}
    \mathbf{M}_{\gamma } =\begin{bmatrix}
\boldsymbol{\gamma }^{\text{H}} & 0 & 0 & 0\\
0 & \boldsymbol{\gamma }^{\eta } & 0 & 0\\
0 & 0 & \boldsymbol{\gamma }^{\eta } & 0\\
0 & 0 & 0 & \boldsymbol{\gamma }^{\text{H}}
\end{bmatrix} ,\ \mathbf{M}_{g} =\begin{bmatrix}
\mathbf{g}^{\text{H}} & 0 & 0 & 0\\
0 & \mathbf{g}^{\eta } & 0 & 0\\
0 & 0 & -\mathbf{g}^{\eta } & 0\\
0 & 0 & 0 & -\mathbf{g}^{\text{H}}
\end{bmatrix} ,\ \mathbf{M}_{1} =\begin{bmatrix}
0 & \mathbf{P} & 0 & 0\\
0 & 0 & \mathbf{R} & 0\\
0 & 0 & 0 & \mathbf{P}^{\dagger }\\
\mathbf{S} & 0 & 0 & 0
\end{bmatrix} ,\ \mathbf{M}_{2} =\begin{bmatrix}
0 & 0 & \boldsymbol{\mu } & 0\\
0 & 0 & 0 & \boldsymbol{\mu }^{\dagger }\\
\boldsymbol{\delta }^{\dagger } & 0 & 0 & 0\\
0 & \boldsymbol{\delta } & 0 & 0
\end{bmatrix}
\end{align}
\begin{align}
    \mathbf{N}_{\gamma } =\begin{bmatrix}
\mathbf{h}_{0} & 0 & 0 & 0\\
0 & \boldsymbol{\eta }_{0} & 0 & 0\\
0 & 0 & \boldsymbol{\eta }_{0} & 0\\
0 & 0 & 0 & \mathbf{h}_{0}
\end{bmatrix} ,\ \mathbf{N} '_{\gamma } =\begin{bmatrix}
\mathbf{h}_{1} & 0 & 0 & 0\\
0 & \boldsymbol{\eta }_{1} & 0 & 0\\
0 & 0 & \boldsymbol{\eta }_{1} & 0\\
0 & 0 & 0 & \mathbf{h}_{1}
\end{bmatrix} ,\ \mathbf{N}_{g} =\begin{bmatrix}
0 & \mathbf{g}_{\parallel }^{X} & 0 & 0\\
0 & 0 & \mathbf{g}_{\parallel }^{\eta } & 0\\
0 & 0 & 0 & \mathbf{g}_{\parallel }^{X\dagger }\\
\mathbf{g}_{\parallel }^{\text{H}} & 0 & 0 & 0
\end{bmatrix} ,\ \mathbf{N}_{1} =\begin{bmatrix}
0 & \mathbf{p} & 0 & 0\\
0 & 0 & \mathbf{r} & 0\\
0 & 0 & 0 & \mathbf{p}^{\dagger }\\
0 & 0 & 0 & 0
\end{bmatrix},
\end{align}
where $\mathbf{E}^{\text{H}}$ and $\mathbf{E}^{\eta}$ are diagonal matrices containing the eigenenergies of the subbands. Without magnetic field, the full Hamiltonian is:
\begin{align}
    \begin{split}
        \mathbf{H}_{0} =\mathbf{E}_{0} +\alpha_{0}\mathbf{M}_{\gamma } k_{\parallel }^{2} +i\alpha _{0}\left(\mathbf{M}_{1} k_{-} -\mathbf{M}_{1}^{\dagger } k_{+}\right) +\alpha _{0}\left(\mathbf{M}_{2} k_{-}^{2} +\mathbf{M}_{2}^{\dagger } k_{+}^{2}\right)
    \end{split}
\end{align}

With $\rho_\pm=(x\pm iy)/2$, the full QW Hamiltonian under magnetic field can be written:
\begin{equation}
\begin{split}
\label{eq:QWprojected}
    \mathbf{H}_\text{QW} =\mathbf{H}_{0}+\frac{\alpha _{0}}{2l_{B}^{2}}\Bigl\{&{\displaystyle \cos \theta \mathbf{L}_{2\perp } +\frac{{\displaystyle \cos^{2} \theta }}{2l_{B}^{2}}\mathbf{L}_{4\perp }} +\frac{\sin^{2} \theta }{2l_{B}^{2}}\mathbf{L}_{4\parallel } +\sin \theta \Bigl[ e^{-i\phi }\mathbf{L}_{2\parallel } \\
    &+\frac{\sin \theta }{2l_{B}^{2}} e^{-2i\phi }\mathbf{L} \prime _{4\parallel } +{\displaystyle \frac{\cos \theta }{2l_{B}^{2}}} e^{-i\phi }\mathbf{L}_{4\times } +\text{h.c.}\Bigr] \Bigr\}
\end{split}
\end{equation}
\begin{align}
&\mathbf{L}_{2\perp } ={\displaystyle \mathbf{M}_{g} +i\mathbf{M}_{\gamma }(\{\rho _{+} ,k_{-}\} -\{\rho _{-} ,k_{+}\})} +2\mathbf{M}_{1} \rho _{-} +2\mathbf{M}_{1}^{\dagger } \rho _{+} -4{\displaystyle i\mathbf{M}_{2} \rho _{-} k_{-}} +4{\displaystyle i\mathbf{M}_{2}^{\dagger } \rho _{+} k_{+}}\\
&\mathbf{L}_{2\parallel } =\mathbf{N}_{g} -2i\mathbf{N} \prime _{\gamma } \rho _{+} -2\mathbf{N}_{1}\{\rho _{+} ,{\displaystyle k_{-}}\} +4\mathbf{N}_{1}^{\dagger } \rho _{+}{\displaystyle k}_{+}\\
&\mathbf{L}_{4\times } =8i\left(\mathbf{N}_{1} \rho _{+} \rho _{-} +\mathbf{N}_{1}^{\dagger } \rho _{+} \rho _{+}\right)\\
&\mathbf{L}_{4\perp } ={\displaystyle 4\mathbf{M}_{\gamma } \rho _{+} \rho _{-} -} 4\mathbf{M}_{2} \rho _{-}^{2}{\displaystyle -} 4\mathbf{M}_{2}^{\dagger } \rho _{+}^{2}\\
&\mathbf{L}_{4\parallel } =8\mathbf{N}_{\gamma } \rho _{+} \rho _{-}\\
&\mathbf{L} \prime _{4\parallel } =-4\mathbf{N}_{\gamma } \rho _{+}^{2}
\end{align}

\section{Effective parameters}
\label{sec:EffectiveParameters}

We give here the effective parameters of a given HH subband $n$. The effective parameters are obtained by performing a SWT on an Hamiltonian that has been projected onto the subband edge basis. In the case of an out-of-plane magnetic field, the result of the projection using the gauge of Eq. (\ref{eq:gaugeBperp}) is equivalent to setting $\theta=0$ in Eq. (\ref{eq:QWprojected}). First order perturbation directly yields linear Rashba $\beta_1=(\mathbf{S})_{nn}$.
Second order perturbation is sufficient to fully represent the parabolic effective mass $\gamma$ and the effective g-factor $\tilde{g}$:
\begin{align}
\gamma =\left( \gamma^{H}\right)_{nn} +\alpha _{0}\left(\sum _{l\neq n}\frac{P_{nl} P_{ln}^{\dagger }}{E_{n}^{H} -E_{l}^{\eta }} +\sum_{l\neq n}\frac{S_{nl} S_{ln}}{E_{n}^{H} -E_{l}^{H}}\right)\\
g^{\mathrm{QW}}_\perp =\left( g^{H}_\perp\right)_{nn} +2\alpha _{0}\left(\sum_{l\neq n}\frac{P_{nl} P_{ln}^{\dagger }}{E_{n}^{H} -E_{l}^{\eta }} -\sum_{l\neq n}\frac{S_{nl} S_{ln}}{E_{n}^{H} -E_{l}^{H}}\right)
\end{align}
To fully represent the cubic Rashba terms $\beta_2$ and $\beta_3$, third order perturbation is necessary and yield:
\begin{align}
\beta _{2} &=\ 2\alpha _{0}^{2}\mathcal{R}\left\{\sum _{l\neq n}\frac{P_{nl} \mu _{ln}^{\dagger }}{E_{n}^{H} -E_{l}^{\eta }}\right\} -\alpha _{0}^{3}\sum _{n'\neq n}\sum_{l\neq n}\frac{P_{nn'} R_{n'l} P_{nl}^{\dagger }}{\left( E_{n}^{H} -E_{n'}^{\eta }\right)\left( E_{n}^{H} -E_{l}^{\eta }\right)}\\
\beta _{3} &=\ 2\alpha _{0}^{2}\mathcal{R}\left\{\sum_{l\neq n}\frac{\delta _{nl} P_{ln}^{\dagger }}{E_{n}^{H} -E_{l}^{\eta }} -\sum_{l\neq n}\frac{\gamma _{nl}^{H} S_{ln}}{E_{n}^{H} -E_{l}^{H}}\right\} 
- \alpha _{0}^{3}\Biggl(\sum _{n'\neq n}\sum_{l\neq n}\frac{S_{nn'} P_{n'l} P_{nl}^{\dagger }}{\left( E_{n}^{H} -E_{n'}^{H }\right)\left( E_{n}^{H} -E_{l}^{\eta }\right)} \\
&+\sum _{n'\neq n}\sum_{l\neq n}\frac{P_{nn'} P_{ln'}^{\dagger } S_{ln}}{\left( E_{n}^{H} -E_{n'}^{\eta }\right)\left( E_{n}^{H} -E_{l}^{\eta }\right)} +\sum _{n'\neq n}\sum_{l\neq n}\frac{S_{nn'} S_{n'l} S_{ln}}{\left( E_{n}^{H} -E_{n'}^{H }\right)\left( E_{n}^{H} -E_{l}^{H}\right)} \\
&-(S)_{nn}\left[\sum_{l\neq n}\frac{{P}_{nl}P^\dagger_{nl}}{(E_n^H-E_l^\eta)^2}+\sum_{l\neq n}\frac{{S}_{nl}S_{ln}}{(E_n^H-E_l^H)^2}\right]\Biggl)
\end{align}
For an in-plane magnetic field, repeating the process of projecting onto the subband edge basis and performing a SWT using instead the gauge presented in Eq. (\ref{eq:gaugeBparr}) leads to:
\begin{equation}
    g^{\mathrm{QW}}_\parallel = 3i\bra{h} [ z q,k_{z}]\ket{h}
\end{equation}
As a sidenote, using a different gauge to obtain a value of $\tilde{g}_\parallel$ is preferable because projecting the full Hamiltonian defined by Eq. ($\ref{eq:QWprojected}$) yields a $\tilde{g}_\parallel$ that will depend on $x$ and $y$, both of which are not well defined in the case of a QW.  

\section{ Hole-Phonon interaction matrix}
\label{sec:relaxationDerivation}
The transition rate from Fermi's golden rule is \cite{li_2020}:
\begin{equation}
    \Gamma_{i f}=\frac{2 \pi \mathcal{V}}{\hbar} \sum_\alpha \int \frac{d^3 q}{8 \pi^3}\left|\left\langle f\left|\mathbf{H}_{\text{H-ph}}\right| i\right\rangle\right|^2 \delta\left(\hbar \omega-\hbar \omega_{\alpha q}\right),
\end{equation}
where $\mathcal{V}$ is the volume of the system, $\hbar\omega_{\alpha q}$ is the energy of the phonon, $\hbar \omega$ the energy difference between the final $|f\rangle$ and initial $|i\rangle$ state of the QD-Thermal phonon bath system, and $\alpha=$\{LA,TA1,TA2\} is the branch index. For absorption, we have $\ket{i} =\ket{\mathbb{0}}\ket{TP} ;\ \ket{f} =b_{\alpha \mathbf{q}}\ket{\mathbb{1}}\ket{TP} /\sqrt{N_{\alpha \mathbf{q}}}$. For emission, $\ket{i} =\ket{\mathbb{1}}\ket{TP} ;\ \ket{f} =b_{a,-\mathbf{q}}^{\dagger }\ket{\mathbb{0}}\ket{TP} /\sqrt{N_{\alpha ,-\mathbf{q}} +1}$. $\ket{TP}$ is the thermal phonon bath with $b_{a\mathbf{q}}^{\dagger } b_{a\mathbf{q}}\ket{TP} =N_{a\mathbf{q}}\ket{TP}$, with occupation number $N_{\alpha q} =1/\left( e^{\hbar \omega _{\alpha q} /k_{B} T} -1\right)$.  The phonon speed is $v_{\mathrm{LA}}=\sqrt{c_{11}/\rho}$ and $v_{\mathrm{TA}}=\sqrt{c_{44}/\rho}$, with $\rho$ the material density. To eliminate the delta function, $q$ is set to $q_\alpha = \omega/v_\alpha$:

\begin{equation}
    \Gamma_{i f}=\frac{\mathcal{V}}{(2 \pi)^2}\left(\frac{\omega}{\hbar}\right)^2 \sum_\alpha \frac{1}{v_\alpha^3} \int \sin \theta d \theta d \phi\left|\left\langle f\left|\mathbf{H}_{\text{H-ph}} \right| i\right\rangle\right|^2
    \label{eq:relax_2}
\end{equation}
The hole-phonon interaction Hamiltonian $\mathbf{H}_{\text{H-ph}}$ is given by (\ref{eq:kpStrain}), with the strain tensor for each polarisation:
\noindent\begin{minipage}{.5\linewidth}
\begin{flalign}
    \varepsilon_{\alpha\boldsymbol{q}} = i q \sqrt{\frac{\hbar}{2\rho \mathcal{V}v_\alpha}}\mathrm{e}^{i\boldsymbol{q}\cdot\boldsymbol{r}}\left(b_{\alpha\boldsymbol{q}}+b_{\alpha-\boldsymbol{q}}^\dagger\right)\epsilon_{\alpha \boldsymbol{q}}
\end{flalign}
\end{minipage}%
\begin{minipage}{.5\linewidth}
\begin{equation}
    (\epsilon_{\alpha\boldsymbol{q}})_{i,j} = \begin{cases}
        \hat{c}_i\hat{q}_i & i=j\\
        1/2(\hat{c}_i \hat{q}_j + \hat{c}_j \hat{q}_i) & i \neq j
    \end{cases}
\end{equation}
\end{minipage}
\noindent
The deformation potential for each polarization are given by:
\begin{equation}
\begin{split}
&\epsilon _{\mathrm{LA}}=\frac{1}{2}\left[\begin{array}{ c c c }
2\sin^{2} \theta \cos^{2} \varphi  & \sin^{2} \theta \sin 2\varphi  & \sin 2\theta \cos \varphi \\
\sin^{2} \theta \sin 2\varphi  & 2\sin^{2} \theta \sin^{2} \varphi  & \sin 2\theta \sin \varphi \\
\sin 2\theta \cos \varphi  & \sin 2\theta \sin \varphi  & 2\cos^{2} \theta 
\end{array}\right],
\epsilon _{\mathrm{TA1}}=\frac{1}{2}\left[\begin{array}{ c c c }
\sin 2\theta \cos^{2} \varphi  & \frac{1}{2}\sin 2\theta \sin 2\varphi  & \cos 2\theta \cos \varphi \\
\frac{1}{2}\sin 2\theta \sin 2\varphi  & \sin 2\theta \sin^{2} \varphi  & \cos 2\theta \sin \varphi \\
\cos 2\theta \cos \varphi  & \cos 2\theta \sin \varphi  & -\sin 2\theta 
\end{array}\right], \\
&\epsilon _{\mathrm{TA2}}=\frac{1}{2}\left[\begin{array}{ c c c }
-\sin \theta \sin 2\varphi  & \sin \theta \cos 2\varphi  & -\cos \theta \sin \varphi \\
\sin \theta \cos 2\varphi  & \sin \theta \sin 2\varphi  & \cos \theta \cos \varphi \\
-\cos \theta \sin \varphi  & \cos \theta \cos \varphi  & 0
\end{array}\right] 
\end{split}
\end{equation}

To project onto the eigenstates of the QW, we separate the $z$-dependant part of the strain tensor:
\begin{equation}
\begin{split}
    \mathbf{H}_{\text{H-ph}}(\varepsilon_{\alpha \boldsymbol{q}}) &=  i q \sqrt{\frac{\hbar}{2\rho \mathcal{V}v_\alpha}}\mathrm{e}^{iq(x\sin \theta \cos \phi +y \sin \theta \sin \phi )}\left(b_{\alpha\boldsymbol{q}}+b_{\alpha-\boldsymbol{q}}^\dagger\right)
    \mathbf{H}_{\text{H-ph}} \left(e^{izq_{\alpha }\cos \theta } \epsilon_{\alpha \boldsymbol{q}}\right) \\
&=  i q \sqrt{\frac{\hbar}{2\rho \mathcal{V}v_\alpha}} \left(b_{\alpha\boldsymbol{q}}+b_{\alpha-\boldsymbol{q}}^\dagger\right)  e^{i q \mathbf{R}_{xy} }\mathbf{W}_z(\epsilon_{\alpha \boldsymbol{q}})
\label{eq:Hinteraction}
\end{split}
\end{equation}

By defining the material parameters $X^\prime=e^{i z q_\alpha \cos \theta }X$, the out-of-plane hole-phonon interaction operator $\mathbf{W}_z$ in the $\{H,\eta\}$ basis can be written:

\begin{align}
\mathbf{W}_{z}^{\text{H} +\text{H} +} &=\bra{h} a^\prime _{v}\ket{h}(\text{Tr}\{\epsilon\}) +\frac{1}{2}\bra{h} b^\prime \ket{h}( \epsilon_{xx} +\epsilon_{yy} -2\epsilon_{zz})\\
\mathbf{W}_{z}^{\text{H} +\eta +} &=i\bra{h} d^\prime \ket{-}( \epsilon_{yz} +i\epsilon_{zx})\\
\mathbf{W}_{z}^{\text{H} +\eta -} &=\bra{h} d^\prime \ket{+} i\epsilon_{xy} -\frac{\sqrt{3}}{2}\bra{h} b^\prime \ket{+}( \epsilon_{xx} -\epsilon_{yy})\\
\mathbf{W}_{z}^{\text{H} +\text{H} -} &=0\\
\mathbf{W}_{z}^{\eta +\text{H} +} &=-i\bra{-} d^\prime \ket{h}( \epsilon_{yz} -i\epsilon_{zx})\\
\mathbf{W}_{z}^{\eta +\eta +} &=\left[\bra{c} a^\prime _{c}\ket{c} +\frac{1}{3}\left(\bra{+} a^\prime _{v}\ket{+} +2\bra{-} a^\prime _{v}\ket{-}\right)\right]( \text{Tr}\{\epsilon\}) +\frac{1}{6}\left(\bra{+} b^\prime \ket{+} -4\bra{-} b^\prime \ket{-}\right)( \epsilon_{xx} +\epsilon_{yy} -2\epsilon_{zz})\\
\mathbf{W}_{z}^{\eta +\eta -} &=\frac{i}{\sqrt{3}}\left(\bra{+} d^\prime \ket{-} -\bra{-} d^\prime \ket{+}\right)( \epsilon_{yz} +i\epsilon_{zx})\\
\mathbf{W}_{z}^{\eta +\text{H} -} &=\bra{+} d^\prime \ket{h} i\epsilon_{xy} -\frac{\sqrt{3}}{2}\bra{+} b^\prime \ket{h}( \epsilon_{xx} -\epsilon_{yy})\\
\mathbf{W}_{z}^{\eta -\text{H} +} &=\bra{+} d^\prime \ket{h}( -i\epsilon_{xy}) -\frac{\sqrt{3}}{2}\bra{+} b^\prime \ket{h}( \epsilon_{xx} -\epsilon_{yy})\\
\mathbf{W}_{z}^{\eta -\eta +} &=\frac{i}{\sqrt{3}}\left(\bra{+} d^\prime \ket{-} -\bra{-} d^\prime \ket{+}\right)( \epsilon_{yz} -i\epsilon_{zx})\\
\mathbf{W}_{z}^{\eta -\eta -} &=\mathbf{W}_{\alpha q}^{\eta +\eta +}\\
\mathbf{W}_{z}^{\eta -\text{H} -} &=-i\bra{-} d^\prime \ket{h}( \epsilon_{yz} +i\epsilon_{zx})\\
\mathbf{W}_{z}^{\text{H} -\text{H} +} &=0\\
\mathbf{W}_{z}^{\text{H} -\eta +} &=\bra{h} d^\prime \ket{+}( -i\epsilon_{xy}) -\frac{\sqrt{3}}{2}\bra{h} b^\prime \ket{+}( \epsilon_{xx} -\epsilon_{yy})\\
\mathbf{W}_{z}^{\text{H} -\eta -} &=i\bra{h} d^\prime \ket{-}( \epsilon_{yz} -i\epsilon_{zx})\\
\mathbf{W}_{z}^{\text{H} -\text{H} -} &=\mathbf{W}_{\alpha q}^{\text{H} +\text{H} +}
\end{align}
The relaxation rate is thus:
\begin{equation}
\begin{split}
        \Gamma = \frac{\omega^3}{8 \pi^2 \hbar \rho} \text{coth}\left(\frac{\hbar \omega}{2 k_B T}\right) \sum_\alpha \frac{1}{v_\alpha^5} \int_0^\pi d\theta \sin \theta \int_0^{2\pi} d \varphi |\langle \mathbb{0}| \mathrm{e}^{i q_\alpha \boldsymbol{r}} \mathbf{W}_z (\epsilon_{\alpha }) |\mathbb{1}\rangle |^2,
    \end{split}
\end{equation}
\newpage
\end{widetext}

\section{Material parameters}

The alloy parameters are obtained from a linear interpolation between the parameters of pure Ge and pure Sn, including a bowing parameter when necessary:
\begin{equation}
    A(x) = (1-x)A^{\mathrm{Ge}}+x A^{\mathrm{Sn}} - b_A x(1-x)
\end{equation}
Luttinger parameters are interpolated from the value of pure Ge and Ge$_{80}$Sn$_{20}$:
\begin{equation}
    \gamma(x) = \left(1-\frac{x}{0.2}\right)\gamma^{\mathrm{Ge}}+\frac{x}{0.2} \gamma^{\mathrm{Ge_{0.80}Sn_{0.20}}}- b_\gamma \frac{x}{0.2}\left(1-\frac{x}{0.2} \right)
\end{equation}
The parameters used for the simulations are presented in table \ref{tab:parametrisationKP}.

\begin{table}
\centering
\caption{Parameters used in $k\cdot p$ calculations}
   \label{tab:parametrisationKP}
 \begin{threeparttable}
    \begin{tabular}{l c c c}
    \hline
    Lattice constant  & Ge  & Sn  & $b$ \\
    a(T=0K) (A)       & 0.5652$^\mathrm{a}$ & 0.6480$^\mathrm{b}$  & -0.0083$^\mathrm{g}$ \\
    \hline
    Energy Gaps (meV) & Ge & Sn & $b$ \\
    $E^\Gamma$ & 898.1$^\mathrm{b}$ & -390$^\mathrm{d}$& 2460$^\mathrm{i}$ \\
       $E^L$ & 785$^\mathrm{c}$ & 100$^\mathrm{g}$ & 1230$^\mathrm{h}$\\
       $\Delta$ & 289$^\mathrm{b}$ & 600$^\mathrm{d}$ & -100$^\mathrm{g}$ \\
       $E_{v,\mathrm{avg}}$ & 0 & 690$^\mathrm{h}$ &  \\ 
    \hline
    Luttinger parameters & Ge &Ge$_{0.80}$Sn$_{0.20}$& $b$ \\
    $\gamma^L_1$ & 13.37$^\mathrm{d}$   & 29.21$^\mathrm{j}$ & 20.34$^\mathrm{j}$ \\
    $\gamma^L_2$ & 4.24$^\mathrm{d}$    & 12.24$^\mathrm{j}$ & 9.67$^\mathrm{j}$ \\
    $\gamma^L_3$ & 5.68$^\mathrm{d}$    & 13.74$^\mathrm{j}$ & 9.82$^\mathrm{j}$ \\
\hline
Other band parameters & Ge & Sn &  \\
    $\kappa$ & 3.41$^\mathrm{d}$     &  -11.84$^{\mathrm{d}}$ & \\
    $ q$ & 0.3$^\mathrm{d}$ & 0.06$^\mathrm{d}$  \\
    $g^*$ & -2.86$^\mathrm{d}$ & 84.4$^\mathrm{d}$&  \\
    $m^* (m_0)$ & 0.038$^\mathrm{d}$ &  -0.058$^\mathrm{d}$ & \\
\hline
Deformation potentials (eV) & Ge & Sn & \\ 
       $\alpha^\Gamma$ & -8.24$^\mathrm{e}$ & -6$^\mathrm{h}$ & \\
       $\alpha^L$ & -1.54$^\mathrm{e}$  & -2.14$^\mathrm{h}$&\\
       $\alpha^v$ &  1.24$^\mathrm{e}$ & 1.58$^\mathrm{h}$ &\\
       $b$ & -2.86$^\mathrm{f}$ & -2.7$^\mathrm{h}$  &\\
\hline
Elastic constants (GPa) & Ge & Sn & \\
       $C_{11}$ & 128.53$^\mathrm{b}$ & 69$^\mathrm{h}$ & \\
       $C_{12}$ & 48.28$^\mathrm{b}$ & 29.3$^\mathrm{h}$& \\
       $C_{44}$ & 66.8$^\mathrm{b}$ & 36.2$^\mathrm{h}$& \\
    \hline
Density (g/cm$^3$) & Ge & Sn \\
$\rho$ & 5.323$^{\mathrm{b}}$ & 7.285$^{\mathrm{b}}$  \\
\hline
    \end{tabular}
    \begin{tablenotes}
      \small
      \item a. ref \cite{reeber1996thermal}
        b. ref \cite{madelung2012semiconductors}
        c. ref \cite{weber1989near}
        d. ref \cite{lawaetz1971valence}
        e. ref \cite{van1989band}
        f. ref \cite{van1986theoretical}
        g. ref \cite{polak2017electronic}
        h. ref \cite{chang2010strain}
        i. ref \cite{bertrand2019experimental}
        j. ref \cite{lu2012electronic}
    \end{tablenotes}
  \end{threeparttable}
\end{table}

\bibliography{apssamp}

\end{document}